# A Holistic Approach to Securing Web Applications

Srdjan Stanković, Dejan Simić

**Abstract**—Protection of Web applications is an activity that requires constant monitoring of security threats as well as looking for solutions in this field. Since protection has moved from the lower layers of OSI models to the application layer and having in mind the fact that 75% of all the attacks are performed at the application layer, special attention should be paid to the application layer. It is possible to improve protection of Web application on the level of the system architecture by introducing new components which will realize protection on higher levels of OSI models. This paper deals with Intrusion Detection Systems, Intrusion Prevention Systems, Web Application Firewall and gives a holistic approach to securing Web applications using aforementioned components.

——————————— ◆ ———————————

## 1 INTRODUCTION

Nowadays Web applications are dominant technology in various activities connected with the Internet. Their faster and faster expansion has been stimulated by the services which provide conditions for the users to exchange and update the information, regardless of software platforms. Web applications usually consist of a program code which is placed and executed by the server side, so they have interaction with database and other sources of dynamic content. The fact that they are spread so much is the reason why Web applications have become critical points.

There are many fields in IT security nowadays which are focused on different security aspects, starting from the lowest layers of OSI models up to the application layer. Since security of the lower layers has been improved a lot, the attackers have become interested in the highest layers of OSI model. They mostly try to find the ways to enter the systems via the application layer. The application layers are extremely exposed when they are on the Internet and used in the form of Web application. Researches have shown that 75% of all attacks take place on the application layer [1].

Large number of the Internet attacks happening today is directed toward exploitation of individuals and organizations in order to earn money and this often causes financial losses. One of the most serious threats to the Internet is presence of large number of compromised computers controlled by one or a few hackers and used for different types of attack. Network of such computers is known as Botnet. Botnets are used for different types of attack, starting from Distributed Denial-of-Service (DDoS), sending of unwanted e-mail messages (SPAM) up to spreading of malicious programs etc [2], [3], [4].

There are several types of approaches to Web application protection: use of appropriate protection strategy [5], security patterns [6], Access Control Systems Combining [7], etc. This paper approaches protection of Web applications from the standpoint of architectural system. It is possible to raise level of Web applications protection to higher level by introducing appropriate components into the system architecture which carries out protection at higher levels of OSI models. The second chapter of this paper deals with the problem description, the third one describes Intrusion Detection Systems and the fourth chapter describes Intrusion Prevention Systems. The fifth chapter describes Web Application Firewalls. The sixth one deals with description of a comprehensive approach to protection of Web application on the level of architecture by use of already mentioned components and they are also mutually compared starting from several aspects.

## 2 PROBLEM STATEMENT

Nowadays Web applications exist in the environment which is totally different from the one that was present in the time when Internet communication was established. Web sites moved from the static HTML pages to dynamic and interactive Web applications. Interactive Web applications have made the Internet universal media and, on the other side, they have brought a new level of security risk.

Web applications are very popular since they are present everywhere on the Internet. Access to the clients' information, financial data and health records are just a few examples of the services which Web applications can offer. In most cases these applications access databases in order to generate dynamic contents. Web systems designed without or with insufficient level of protection can cause loss of data, loss of access, loss of confidence and privacy.

Botnets are one of the most serious threats to Web applications. Botnets or nets of infected computers with one or more controllers can cause a lot of problems in functioning of Web systems/applications [8].

————————————————

- *Srdjan Stanković, Ministry of Defense, Republic of Serbia, Nemanjina 15, 11000 Belgrade, Serbia.*
- *Dejan Simić, Faculty of Organizational Sciences, University of Belgrade, Jove Ilića 154, 11000 Belgrade, Serbia.*

An example of architecture for Web application (Figure 1) presents one of the ways in which database server and Web server can be connected with the end user of Web application. Implementation of firewall is standard first step in the system realization. Firewalls protect against many attacks from the network and system infrastructure.

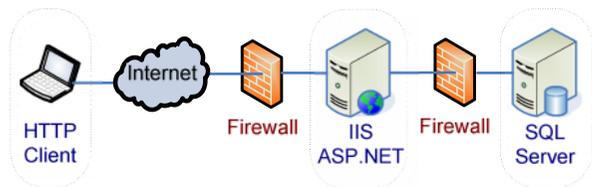

Fig. 1. An Example of Architecture for Web applications

Besides, some firewalls enable filtering of the contents such as malicious Java and Active-X applications. However, Firewall cannot do too much in protection against malicious incoming demands sent to the application. Besides firewall, proper configuration of active network equipment (Routers and Switches), maintaining/updating of appropriate anti-virus protection as well as permanent up-dating of the system software are compulsory activities of the system protection [9]. However, all these activities are not sufficient in modern time to protect the system against malicious actions which the attackers try to apply against Web applications. It is necessary to upgrade the system architecture in order to improve level of protection.

## 3 INTRUSION DETECTION SYSTEM

Intrusion Detection System (IDS) is a software and/or hardware platform which collects information about the network traffic from a number of network and computer sources and analyses them in order to detect possible improper activities and misusages (Figure 2). IDS follows and analyses network traffic but it doesn't do anything to prevent unwanted traffic [10]. If unwanted traffic is noticed, in most cases the persons in charge of the traffic are informed.

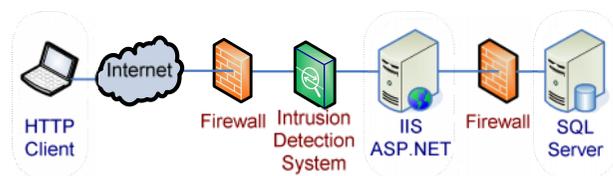

Fig. 2. An Improved Architecture with IDS for Web applications

There are several kinds of IDS and they are as follows [11]:
- Network Intrusion Detection System (NIDS) is a device intended for collecting and analyzing of network traffic of the devices which are connected to Hubs or Switches and the device is not necessarily produced or directed toward the computer IDS is installed on.
- Protocol-based Intrusion Detection System (PIDS) consists of a system or agent which is usually placed directly in front of server and which follows and analyses communication protocol between the mutually connected devices (user of the computer or system) and server. It usually understands and follows HTTP protocol for Web servers in order to protect them.
- Application Protocol-based Intrusion Detection System (APIDS) consists of a system or an agent which is usually part of a group of servers and which follows and analyses communication of specific applicable protocol.
- Host-based Intrusion Detection System (HIDS) consists of an agent on the host computer which identifies raids via analysis of the system of calls (invitations), applications logs and other activities on the host computer.
- Hybrid Intrusion Detection System is combination of two or more approaches. The information collected by the host's agent and the information from the network give entire review of whole network of the activities.

Work of IDS tools is based on checking of different logs and content of various network packets. Both types of data contain information about the user's activities so analysis of their content enables identification of unauthorized activities. There are two methods that are usually applied nowadays [12]:
- Signature based – contents of logs and/or network packets are sent to search program and they are compared to signatures of attacks defined in advance.
- Statistical anomaly – Anomalies which are reported as attack against the system are recognized based on the profile of habitual work of the system made by collecting and statistic processing of the data.

Advantage of IDSs use is that they do not slow down network traffic, but there are numerous disadvantages of IDSs such as:
- Detection without prevention – ISDs cannot stop or slow down active network attacks.
- Time discrepancy between attack and detection– since there is large number of records which must be analyzed by IDSs as well as because of possible need for additional checking of the administrator's positives about attack, IDSs slowly responses to the attacks against computer and network resources.
- False Positive – While reviewing content of network packets and daily logs, IDSs often qualify the proper traffic as improper one and reverse; this happens because of too many generalized definitions of signatures of some attacks.
- Large number of daily logs about attacks - since IDS is not able to stop an attack, it generates as many warnings as there are daily logs about the very attacks. Since legal traffic is sometimes without justification defined as an attack, number of warnings sent to the administrators additionally increases.
- Absence of detection of new kinds of attacks – since IDSs mostly compare contents of daily logs and network packets with defined signatures for attacks, new attacks are not registered in database and IDS cannot recognize them. Detection of unknown attacks is possible only in case IDS is based on the method of the statistic anomalies recognition.

Because of the above mentioned disadvantages, it is necessary to move the level of security from detection of malicious activities to their prevention. That's why Intrusion Prevention System has been developed and it primarily enables prevention actions and avoiding of harm caused by the attacks.

## 4 INTRUSION PREVENTION SYSTEM

Intrusion Prevention System (IPS) came out of IDS, and it is a device which monitors the network activity in order to detect malicious or unwanted activities in realistic time, to block them and prevent them from further activity (Figure 3) [5].

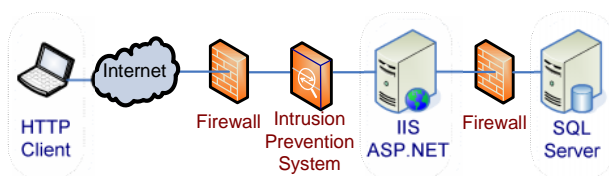

Fig. 3. An Improved Architecture with IPS for Web applications

IPSs function in such a way that when they detect an attack or unwanted behavior they drop unwanted packets and let the rest of the traffic pass. Although IPSs are similar to Firewalls, their way of approaching the network and security system is totally different [13]. Because of their similarity, IPSs and Firewalls are often confused and treated as the same thing.

IPS is usually designed in such a way that it works in the network in a completely invisible way. IPS can respond to the attack in one of the following ways:
- dropping packets,
- resetting connection,
- generating warning and
- quarantining intruders.

Some IPSs are able to apply the rules of Firewall, but it is often only its advantage but not basic function. Furthermore, IPS technology provides better insight into network traffic by giving the information about too active hosts, bad logons, unsuitable content and many other functions of the network and applicable layers.

Content can be filtered in several ways and, based on the ways, they are classified into several types. The most important types are [14]:
- Content-based one controls content of the packet, looking for characteristic sequence or signature for detection and prevention of already known attacks such as infections caused by worms and Trojans.
- Protocol Analysis one can decode application layer protocols such as HTTP or FTP. After complete decoding of the protocol, IPS analysis engine can detect unusual behavior, i.e. unexpected values in the packets and drop such packets.
- Rate-based ones are primarily intended for prevention Denial of Service (DoS) and Distributed Denial of Service (DDoS) attacks. They monitor and learn normal behavior of the network. They follow the traffic in realistic time and compare it with memorized statistics and, based on this, can recognize abnormal rates of certain types of traffic such as TCP, UDP or ARP packets. The attacks are identified when they exceed the limits which are dynamically adjusted depending on part of the day, day of the week etc., in accordance with memorized statistics.

Since work of IPSs is based on IDS, they have inherited some of their negative phenomena or lacks, most of all False Positives and Absence of new attacks detection.

## 5 WEB APPLICATION FIREWALL

Web Application Firewall (WAF) is a device either server Plug-in or filter which applies a number of rules with HTTP/ HTTPS conversation [15]. WAF realizes proper context of HTML demand and it responses and secures semantic correctness for various objects present on Web application such as different kinds of fields, falling lists, server and client scripts, attached input and output parameters… It can block the attacks for wide variety of server and program platforms as well as database platforms. It can control both incoming and outgoing traffic [16]. Security of Web applications can be significantly improved by upgrading the architecture with Web Application Firewall (Figure 4).

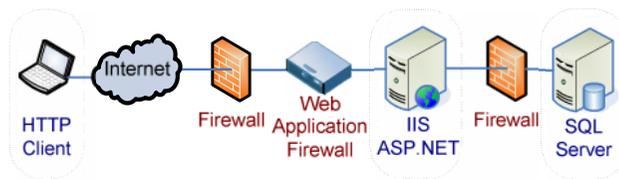

Fig. 4. An Improved Architecture with WAF for Web applications

WAF can work in several different modes. Each mode has its advantages and disadvantages and it is necessary to estimate them properly before they are built into certain system [17]. WAF modes of work are:
- Reverse Proxy – Full Reverse Proxy is the most common mode and it has many functions. The device is placed between Firewall- and Web server; it has public IP address and all incoming traffic comes to the address. After that WAF creates demand towards Web server on behalf of the user browser. This way of work is often necessary for many additional functions which can be provided by WAF because there is possibility of the connection breaking. Disadvantage of the Reverse Proxy mode is that it causes lateness (delay) and some problems can appear with some applications.
- Transparent Proxy – When used as a Transparent Proxy, WAF is placed between Firewall and Web server, like Reverse Proxy, but it doesn't have IP address. This regime doesn't demand any changes in the existing infrastructure, but it cannot provide additional services which can be provided by Reverse Proxy.
- Layer 2 Bridge - WAF is placed between Firewall and

Web server and it functions like a layer 2 switch. This regime provides high level performances and there are not significant changes in the network. However, it doesn't offer advanced services offered by other WAF modes.
- Network Monitor/Out of Band – WAF is not in the same line with Firewall and Web server in this mode, but it monitors the traffic on selected port. This regime is ideal one for WAF testing in certain surroundings without influencing upon the traffic. If needed, WAF can still block the traffic in this regime by sending TCP reset in order to interrupt unwanted traffic.
- Host/Server Based – Host or Server based WAF is a software application which is installed on Web server. Host based WAF doesn't provide any additional functions which can be offered by its network based mate. Host based WAF increases load on Web server, so introducing of this application to already loaded servers should be done very carefully.

WAF often has additional components or they can introduce additional possibilities for improvement of reliability and performances of Web applications. These additional functions can decide if certain WAF will be used in some organization. It is not that all WAFs have additional functions but many of them depend on the applied mode of work. Reverse Proxy usually supports all the following functions [17]:
- Caching – Reduction of load on Web server and increasing of performances by caching the copies of Web content. In this way number of repeated demands to the server relating recently sent pages/content is decreased.
- Compression – In order to provide efficient network traffic, WAF can automatically compromise some Web contents which are later unpacked by the user browser.
- SSL Acceleration – Use of WAF based on hardware SSL description increases speed of SSL processing and decreases load on Web server.
- Load Balancing – Spreading of incoming Web demands to several Web servers in order to improve performance and reliability.
- Connection Pooling – Reduces back end server TCP overhead by allowing multiple requests to use the same back end connection.

If WAF is properly chosen and if the rules for certain user Web application are adjusted, WAF can identify and block many attacks (Cross-site Scripting, SQL Injection…) [18]. Certain effort is necessary when the rules are updated having in mind the fact that when Web applications are changed rules for work of WAF also must be changed.

## 6 A HOLISTIC APPROACH TO SECURING WEB APPLICATIONS

Measures for the system protection are carried out by applying of appropriate politics which means physical, organizational, software protection measures, etc. Besides, one of very important factors influencing upon security is proper coding of Web application without security failures and using already checked procedures and solutions for the application protection. Protection on the level of architecture, which is the central part of this paper and which is different from standard solutions, is a significant improvement of protection system in Web systems.

There are several kinds of devices for protection of Web applications (IDS, IPS and WAF) and we must be very careful when deciding how to choose the most optimal combination. Since IDS are used only for detection and IPS as their successor has capability to block unwanted traffic, it is clear that IPS is the better choice. If we decide to build in IPS, it is not necessary to build in IDS. In order to achieve better protection of Web applications, WAF should also be built in (besides IPS). In that case protection on the level of the system architecture is overall. When both devices are built in they mutually supplement each other while protecting the system. Appropriate combination of these two devices and their proper adjusting provide successful protection of Web applications against all kinds of attack (Figure 5).

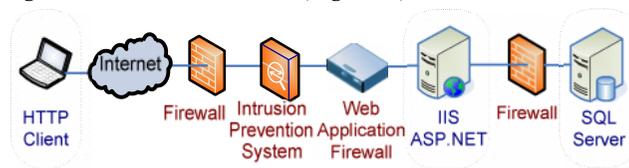

Fig. 5. A Holistic Approach to Securing Web applications

For better understanding activities of Firewall, IDS, IPS and WAF, comparative reviews of technologies [16], differences in architecture [19] and review for several criteria relating introduction and maintenance are given in the tables.

TABLE 1
COMPARATIVE REVIEW OF TECHNOLOGIES

| Technology | Purpose | Way of work | Area of action |
|---|---|---|---|
| Network Firewall | Protection of network protocols | Blocking of TCP and UDP ports | Network protocols |
| Intrusion Detection System | Network monitoring | Traffic scanning | Network protocols Network applications |
| Intrusion Prevention System | Network protection | Traffic scanning, Packet dropping | Network protocols Network applications |
| Web Application Firewall | Protection of HTTP/HTTPS applications | URL normalization, Realization of session state, Realization of application context | HTTP/HTTPS applications |

TABLE 2
REVIEW OF DIFFERENCES IN ARCHITECTURE

| Architecture/design | IDS | Firewall/IPS | WAF |
|---|---|---|---|
| Understands HTTP | Yes | Yes | Yes |
| Understands HTML | No | No | Yes |
| Control of the network access | No | Yes | No |
| Web applications protection | No | No | Yes |
| IP packet control | No | Yes | No |
| Semantic understanding of stream | No | No | Yes |
| Blocks network attacks | No | Yes | No |
| Blocks attacks against application | No | No | Yes |

TABLE 3
INTRODUCING AND MAINTENANCE

| Criterion | Firewall | IDS | IPS | WAF |
|---|---|---|---|---|
| Price | low | medium | high | high |
| Necessary changes in architecture | yes | no | yes | yes |
| The time needed and results | immediately | take time | take time | immediately |
| Maintenance | simple | difficult | difficult | difficult |
| Upgrade | rarely | frequently | frequently | frequently |

## 7 CONCLUSION

There are many security threats for Web applications. Botnets are example for nowadays significant security challenge for the Internet community. Their activities cause a lot of troubles among the Internet users, system administrators, providers etc. Threats imposed by Botnets require efficient protection systems of computers and Web applications.

Protection on the level of the system architecture is very important for protection of Web applications. The paper has shown differences in architecture, comparative review of devices/technologies as well as review for several criteria relating introduction and maintenance of Firewall, Intrusion Detection System, Intrusion Prevention System and Web Application Firewall so that it is possible to choose proper type of Prevention System and Web Application Firewall with appropriate mode which is the most suitable for certain system, depending on the very system architecture when it is developed, upgraded and maintained. The conclusion is that Web applications can be successfully protected against the attacks by upgrading the system architecture with Intrusion Prevention System and Web Application Firewall as described in the paper.

**Srdjan Stanković** is a MSc student at the Faculty of Organizational Sciences, University of Belgrade. He is working at Ministry of Defense as a Software Analyst. His interests are Web application security, security of computer systems, applied information technologies.

**Dejan Simić**, PhD, is a professor at the Faculty of Organizational Sciences, University of Belgrade. He received the B.S. in electrical engineering and the M.S. and the Ph.D. degrees in Computer Science from the University of Belgrade. His main research interests include: security of computer systems, organization and architecture of computer systems and applied information technologies.